\RequirePackage{fix-cm}
\documentclass[twocolumn,epjc3]{svjour3}  
\smartqed  
\RequirePackage{graphicx}
\RequirePackage{mathptmx}      
\usepackage{amsfonts}
\usepackage{amsmath,amssymb}
\RequirePackage{latexsym}
\RequirePackage[numbers,sort&compress]{natbib}
\RequirePackage[colorlinks,citecolor=blue,urlcolor=blue,linkcolor=blue]{hyperref}
\journalname{Eur. Phys. J. C}
\begin{document}

\title{Wave zone in the Ho\v{r}ava-Lifshitz theory at the kinetic-conformal point in the low energy regime}


\author{J. Mestra-P\'aez\thanksref{e1,addr1}
        \and
        J.M. Pe\~na\thanksref{e2,addr1} 
        \and 
        A. Restuccia\thanksref{e3,addr1}
}

\thankstext{e1}{e-mail: jarvin.mestra@ua.cl}
\thankstext{e2}{e-mail:joselen@yahoo.com}
\thankstext{e3}{e-mail:alvaro.restuccia@uantof.cl}


\institute{Departamento de F\'isica, Universidad de Antofagasta, Aptdo 02800, Chile. \label{addr1}
}

\date{Received: date / Accepted: date}

\maketitle

\begin{abstract}
We show that in the Ho\v{r}ava-Lifshitz theory at the  kinetic-conformal point, in the low energy regime, a wave zone for asymptotically flat fields can be consistently defined. In it, the physical degrees of freedom, the transverse traceless tensorial modes, satisfy a linear wave equation. The Newtonian contributions, among which there are terms which manifestly break  the relativistic invariance, are non-trivial but do not obstruct the free propagation (radiation) of the physical degrees of freedom. For an appropriate value of the couplings of the theory, the wave equation becomes the relativistic one in agreement with the propagation of the gravitational radiation in the wave zone of General Relativity. Previously to the wave zone analysis, and in general grounds, we obtain the physical Hamiltonian of the Ho\v{r}ava-Lifshitz theory at the kinetic-conformal point in the constrained submanifold. We determine the canonical physical degrees of freedom in a particular coordinate system. They are well defined fuctions of the transverse-traceless modes of the metric and coincide with them in the wave zone and also at linearized level.

\end{abstract}

\section{Introduction}

The detection of gravitational waves has opened a new era in the study of physics \cite{Miller2019, Nitz2021}.
Multi-messenger astronomy will be decisive in the study of astrophysical and cosmological events
and can lead to the discovery of new phenomena in extreme situations beyond the reach of
experimental tests that we now carry out.

Ho\v{r}ava-Lifshitz gravity is a recent proposal for a candidate to ultraviolet completion of General Relativity (GR) \cite{horava2009,Blas2010}. This theory models the gravity as 4-dimensional differentiable manifold with a foliation-structure of co-dimension one. The foliation-leaves are 3-dimensional Riemannian submanifolds.  
In addition, the time and space scale in different ways, consequently the relativistic symmetry is manifestly broken. The anisotropic scaling allows to include interaction terms with high spatial derivatives in the potential, without breaking the symmetry of the action under diffeomorphisms that preserve the foliation  while keeping the second-order time derivatives of the kinetic term. 
The theory contains several coupling constants. There is only one in the kinetic term of the Ho\v{r}ava-Lifshitz action, it is  dimensionless and plays a relevant role in  the theory. When its value is $\lambda=1/3$, the so-called kinetic-conformal point, the theory propagates, at the linearized level, the same degrees of freedom of linearized General Relativity and with an appropriate choice of coupling parameters, it is consistent with low energy  experiments \cite{EmirGumrukcuoglu2018, BellorinRestucciaSotomayor_2013}.
At linearized level the non-propagating components of the metric and the lapse function become zero in both theories  as a consequence of the constraints.

The original Ho\v{r}ava-Lifshitz gravity \cite{horava2009}  suffers from a strong coupling problem associated to the scalar mode of the theory. An improved formulation was obtained by the inclusion of a new interaction term proposed by Blas, Pujol\`as and Sibiryakov (BPS), quadratic in derivatives and compatible with the foliation preserving diffeomorphisms \cite{Blas2010,Blass2010a} and the corresponding contributions to the potential. This extended formulation is free from the strong coupling problem of the original Ho\v{r}ava-Lifshitz version. It is a weakly coupled anisotropic description of gravity.

The Ho\v{r}ava-Lifshitz gravity for $\lambda=1/3$, with the inclusion of the BPS interaction term, does not propagate, at any energy scale, the scalar mode which occurs for $\lambda \neq 1/3$. Therefore, it does not have any strong coupling problem. The inclusion of the BPS interaction term is essential to have a formulation of the second class constraints in terms of strictly elliptic partial differential equations, which allow a consistent elimination of non-physical modes. Besides, the field equations for the $\lambda=1/3$ model, evaluated at $\alpha=0$ and $\beta=1$ coincide exactly with the General Relativity field equations in the gauge $\pi=0$ (an admissible  gauge condition outside the black hole horizon in GR). This gauge was used in the ADM analysis of the dynamics of GR \cite{ADM2008}. Moreover, for the models with $\lambda \neq 1/3$ the only dependence on $\lambda$ in the Hamiltonian is through the term proportional to $\pi^{2}$ given by

\begin{equation}\frac{N}{\sqrt{g}} \frac{\lambda}{(3\lambda – 1)} \pi^{2}.  \end{equation}

On the other hand, in the Hamiltonian of the $\lambda=1/3$ model there is no $\pi^{2}$ term, since there is a second class constraint $\pi=0$ which arises directly as a primary constraint from the canonical formulation of the theory. The term $\pi^{2}$ could in principle be generated from the renormalization group flow, since it is an admissible term in the Hamiltonian under the foliation preserving diffeomorphisms. However, the second class constraint which has to be imposed at all times prevents its appearance. Consequently, the BPS extension of the $\lambda=1/3$ Ho\v{r}ava-Lifshitz gravity is a viable model, describing anisotropic gravity.

There are then three non projectable viable models of anisotropic gravity: the healthy extension of Ho\v{r}ava-Lifshitz \cite{Blas2010} for $\lambda \neq1/3 $, which describes the propagation of transverse-traceless tensorial modes together with a scalar one, the $U(1)$ symmetric models \cite{Horava2010,WangAnzohongWu2011,daSilva2011,Kluson2011,Linetal2011}   for the projectable models and \cite{ZhuEtal2011} for the non projectable models,   and finally  the $\lambda=1/3$ model of Ho\v{r}ava-Lifshitz gravity, the conformal kinetic model, which with the inclusion of the BPS interaction term, propagates only the transverse-traceless tensorial modes. We will consider in this paper this latest model.
The theory ends up being power counting renormalizable and unitary \cite{Bellorin_Restuccia_2016B,charmousis2009,visser2009,papazoglou2010,orlando2009,shu2009,benedetti2014,contillo2013,d2014asymptotic,d2015covariant, barvinsky2016,wang2017,shin2017,pospelov2012}. 

Beyond the linearized formulation, GR has a well defined wave zone, on which the physical degrees of freedom propagate freely  on a nontrivial Newtonian background. This is a nontrivial property of GR not necessarily valid for nonlinear theories. Hence, we may wonder if such a property is also valid for Ho\v{r}ava-Lifshitz gravity, and if it is the case what are the effects of the anisotropy, that is, the breaking of the Lorentz symmetry, on the wave zone.
Arnowitt, Deser and  Misner (ADM) proved in the 60's that General Relativity  has a well defined wave zone \cite{Arnowitt1961}.  In this space-time region the metric components  $g^{TT}_{ij}$ at order $\mathcal{O}(1/r)$ satisfy the same wave equation as in the linearized  theory around a Minkowski space-time. In the wave zone there also exists a Newtonian background at order $\mathcal{O}(1/r)$ that  does not prevent the $g^{TT}_{ij}$ modes, the physical degrees of freedom,  to propagate as free radiation \cite{Arnowitt1961,ADM2008}. 

We prove in this work that in the $(\lambda=1/3)$-Ho\v{r}ava-Lifshitz gravity at low energies, although the relativistic symmetry is broken, there is a wave zone where the physical degrees of freedom, the transverse traceless tensorial modes, propagate freely satisfying a wave equation, without any interaction with the nontrivial Newtonian background as in GR.
This means that the low energy results obtained for the wave propagation in GR, in the wave zone, are also valid for the Ho\v{r}ava-Lifshitz gravity theory.

We emphasize that although  in the wave zone the dynamical equation satisfied by the physical degrees of freedom is the same as the one obtained from a linearized analysis, in the wave zone there are nontrivial Newtonian components of the metric (background). This Newtonian background, a nonlinear effect of the analysis, determine the energy and momentum of the gravitational wave. This situation  also happens in GR, but with different expressions for the energy and momentum \cite{Blas2011}.

We show, using the local symmetries, that the Ho\v{r}ava-Lifshitz theory, at the kinetic-conformal point, can be canonically reduced to a Hamiltonian formulation in terms of the physical degrees of freedom,  in a particular coordinate frame. The Hamiltonian density explicitly depends on the Lorentz violating term characterized by the coupling parameter $\alpha$. 

In section \ref{Foliation}, we review the Ho\v{r}ava-Lifshitz theory at the kinetic-conformal point. In section \ref{Energy}, we show that there is a Hamiltonian formulation in terms of the physical degrees of freedom. We obtain the Hamiltonian density in a particular coordinate system.  In section \ref{wavezone}, we obtain the formulation of the theory in the wave zone. Finally, in section \ref{Conclusion}, we give the conclusions of our paper.

\section{Foliation, geometry and Ho\v{r}ava-Lifshitz gravity}
\label{Foliation}
Let $M$ be a 4-dimensional differentiable manifold. $M$ has a codimension-one foliation structure $(M,\mathcal{F})$ if  the  maximal atlas $\mathcal{F}\equiv(U_{\mathcal{A}}, \varphi_{\mathcal A})$ i.e $M=\cup \, U_{\mathcal{A}}$, where $U_{\mathcal{A}}$ is a family of open subsets of $M$ and $\varphi_{\mathcal{A}}:U_{\mathcal{A}}\rightarrow D_{\mathcal{A}} \subset \mathbb{R}^{1}\times \mathbb{R}^{3} $ are diffeomorphism such that if $U_{i}\cap U_{j}\neq \emptyset$ the transition of charts is defined by  
\begin{eqnarray}\varphi_{i}\circ \varphi^{-1}_{j}: \varphi_{j}(U_{i}\cap U_{j})\rightarrow \varphi_{i}(U_{i}\cap U_{j}),\\
	(t,x)\rightarrow (\tilde{t}(t),\tilde{x}(t,x)).
\end{eqnarray} 

The couple $(M,\mathcal{F})$ and its equivalents under the diffeomorphisms, that preserve the foliation structure, $\mathcal{F}_{Diff}$,  provide the geometrical structure of the Ho\v{r}ava-Lifshitz theory where space and time scale anisotropically $t\rightarrow b^{z}t$ and $x\rightarrow bx$. 
We remark that $M$ is the disjoint union of 3-dimensional Riemannian manifolds $(\Sigma_{t},g_{ij})$, $	g_{i j}(t,x) = \frac{\partial \tilde{x}^{l}}{\partial x^{i}} \frac{\partial \tilde{x}^{m}}{\partial x^{j}}\tilde{g}_{lm}(\tilde{t},\tilde{x})$, where the following geometric objects compatibles with the foliation structure are introduced: a proper time defined  through the introduction of the lapse $N$ and a shift of the spatial coordinates defined through $N^{i}$   in order to have a contravariant transformation law under $\mathcal{F}_{Diff}$.
\begin{eqnarray}
\label{transformation}
\tilde{N}(\tilde{t}, \tilde{x}) d \tilde{t} = N(t,x) dt,  \\
d\tilde{x}^{i} + \tilde{N}^{i}(\tilde{t}, \tilde{x}) d \tilde{t}= \frac{\partial \tilde{x}^{i}}{\partial x^{j}}[dx^{j} + N^{j}(t,x) dt]\,,
\end{eqnarray}	
we emphasize that there is not a space-time metric on $M$. The metric on the leaves of the foliation $g_{ij}$ and the fields $N$ and $N^{i}$ are used to describe the evolution of the gravitational field. They  scale anisotropically as $ g_{ij} \rightarrow  b^{0} g_{ij}$, $N\rightarrow b^{0} N$ and $N^{i}\rightarrow b^{z-1}N^{i}$.

Taking into account the anisotropic scaling and foliation structure the proposal incorporates terms with high spatial derivatives in the potential without breaking the symmetry under $\mathcal{F}_{Diff}$. The Hamiltonian of the Ho\v{r}ava-Lifshitz gravity theory at the kinetic-conformal point is given by \cite{RestucciaTello2020, BellorinRestucciaTello2018, Restuccia-Tello_2020,BellorinRestucciaSotomayor_2013}

\begin{eqnarray}
\label{hamiltonian_with_surface_terms}
H=\int_{\Sigma_{t}} d^{3}x\bigg\{N\sqrt{g}  \left[ 
	\frac{\pi^{ij}\pi_{ij}}{g}- \mathcal{V}\left(g_{ij},N\right)
	\right]  \nonumber \\
	- N _{j}H^{j}- \sigma P_{N}-\mu \pi \bigg\}+\beta E_{ADM} \,, 
\end{eqnarray}
here $\pi^{ij}$ is the canonical conjugate of $g_{ij}$, $N_{i}\equiv g_{ij}N^{j}$, $\sigma$ and  $\mu$ are Lagrange multipliers. The  surface integral 
\begin{eqnarray}
\label{ADMenergy}
  E_{ADM}&\equiv& \oint_{\partial \Sigma_{t}} \left( \partial_{j}g_{ij}-\partial_{i}g_{jj}\right)dS_{i}\,,
\end{eqnarray}
%
%
is added in order to ensure the Fr\'echet differentiability of the Hamiltonian, see \cite{Regge-Teitelboim1974}
where this idea was introduced and it was shown that $E_{ADM}$ is the ADM energy in GR.
%
The potential, up to terms quadratic on the Riemann tensor and the vector field $a_i$,  is $\mathcal{V}= \mathcal{V}^{(1)}+\mathcal{V}^{(2)}+\mathcal{V}^{(3)}$, with \cite{Bellorin_Restuccia_2016B}
\begin{eqnarray}
	\label{potential1}
	\mathcal{V}^{(1)}&=& \beta R + \alpha a_ia^i, \\
	\label{potential2}
	\mathcal{V}^{(2)}&=& \alpha_{1} R \nabla_i a^i +  \alpha_{2} \nabla_i a_j \nabla^i a^j  +  \beta_{1} R_{ij} R^{ij} + \beta_{2} R^{2}\\
	\label{potential3}
	\mathcal{V}^{(3)}&=& \alpha_{3}\nabla^2 R \nabla_i a^i +  \alpha_{4} \nabla^2 a_i \nabla^2 a^i +  \beta_{3} \nabla_i R_{jk}\nabla^i R^{jk} \nonumber \\&&+ \beta_{4} \nabla_i R \nabla^i R \,,
\end{eqnarray}
where $a_i\equiv \frac{1}{N}\partial_{i} N$, $\nabla_{i} $ represents the  affine  connection, the covariant derivative constructed with the Riemannian metric on the leaves, $\alpha$'s and $\beta$'s are coupling constants.
The potential also contains terms of the same order in the spatial derivatives as the ones explicitly shown, but of cubic order or greater on the Riemann tensor and the vector field $a_i$. Since they do not contribute to the dominant order in the wave zone they have not been presented in (\ref{potential2}) and (\ref{potential3}).
The primary constraints of the theory  are
\begin{eqnarray}
\label{vinculo-momento}
H^{j}= 2\nabla_{i}\pi^{ij}=0,\quad
P_{N}=0,\quad
\pi=g_{ij}\pi^{ij}=0\,.
\end{eqnarray}
%
%
%
%

If we consider only the low energy potential, $\mathcal{V}^{(1)}$, the  time preservation of primary constrains imply the following secondary constraints: 
\begin{eqnarray}
\label{H_P-constraint}
 H_P &\equiv& \frac{3}{2} \frac{1}{\sqrt{g}} \pi^{ij}\pi_{ij}+\frac{1}{2}\sqrt{g}\beta R \nonumber \\&&
+ \sqrt{g}\left(\frac{\alpha}{2}-2\beta \right)a_{i}a^{i}- 2\beta \sqrt{g}\nabla^{i}a_{i}=0, \\
\label{H_N-constraint}
 H_N &\equiv& \frac{1}{\sqrt{g}}\left(\pi^{ij}\pi_{ij}- \beta g R \right)+ \alpha \sqrt{g}a_{i}a^{i} +\nonumber \\&&2\alpha \sqrt{g}\nabla_{i}a^{i}=0 \,,
\end{eqnarray} 
which together with the last two constraints in (\ref{vinculo-momento}) are  second class constraints, while $H^j=0$ is a first class constraint.

The evolution equations are,
\begin{eqnarray}
	\label{gpunto}
	\partial_t {g}_{ij}&=&\frac{2N}{\sqrt{g}}\pi_{ij}+ 2\nabla_{(i}N_{j)}- \mu g_{ij}\,, \\
	\label{pipunto}
	\partial_t{\pi}^{ij}&=&\frac{N}{2}\frac{g^{ij}}{\sqrt{g}}\pi^{kl}\pi_{kl}-  \frac{2N}{\sqrt{g}}\pi^{ik}\pi^{j}{}_{k}+N \sqrt{g}\beta \left(\frac{R}{2} g^{ij}-R^{ij}\right) \nonumber \\
	&&\nonumber \\ &&
	-\alpha N\sqrt{g}\left(a^{i}a^{j}-\frac{1}{2}g^{ij}a_{k}a^{k}\right)- \nabla_{k}\left[2\pi^{k(i} N^{j)}-\pi^{ij} N^{k}\right]\nonumber \\
	&&+\beta\sqrt{g}\left[\nabla^{(i}\nabla^{j)}N-g^{ij}\nabla^{2}N\right]+\mu \pi^{ij}  \,. \qquad
	\end{eqnarray}

\section{Energy in Ho\v{r}ava-Lifshitz gravity}
\label{Energy}
If we calculate the Hamiltonian (5) in the constrained submanifold  it reduces to a surface term. We are going to show in this section that the physical Hamiltonian, that is, the Hamiltonian expressed  solely in terms of the physical degrees of freedom is the mentioned surface term. To do so, we evaluate the Lagrangian on the constrained submanifold and express the kinetic terms solely  in terms of the physical degrees of freedom. From it we obtain the physical Hamiltonian which ends up being the surface term $E$. This is a way to show that the gravitational energy is given by the surface term $E$. An interesting  point in our argument will be that no integration by parts needs to be performed.

In order to identify which  field components  propagate and which are static ones, we will use an orthogonal linear decomposition in transverse  and longitudinal parts, the ADM decomposition \cite{Arnowitt1961,ADM2008}. A symmetric tensor that vanishes at infinity can be expressed as 
\begin{equation}
	\label{T+L}
	f_{ij}=f_{ij}^{TT}+f_{ij}^{T}+2\partial_{(j}f_{i)}\,.
	\end{equation}
	The transverse part $f_{ij}^{T}\equiv\frac{1}{2}[\delta_{ij}f^{T}-\frac{1}{{\Delta}} \partial_{i}\partial_{j}f^{T}]$
	is divergence-free. The  $f_{ij}^{TT}$-part
	is divergence-free and trace-free.   The remaining term  $2\partial_{(j}f_{i)}$ is its longitudinal part. $f^{T}=\delta ^{ij}f_{ij}^{T}$ is the trace of the transverse part of $f_{ij}$.   $\frac{1}{\Delta}$ is the inverse of the flat space Laplacian,  defined on the space of functions which vanish at infinity.

It is known that the Hamiltonian of the Ho\v{r}ava-Lifshitz action can be rewritten as a linear combination of constraints \cite{Bellorin_Restuccia_2016B} provided two surface terms are included:
\begin{equation}
\label{E_boundaries_terms}
E=- \beta \oint_{\partial \Sigma_{t}} dS_i g^T_{,i}  -2 \alpha \oint_{\partial \Sigma_{t}} dS_i N_{,i} \,\,, 
\end{equation}
and that the physical degrees of freedom of the Ho\v{r}ava-Lifshitz gravity at the linearized level  are the TT tensorial modes.
Notice that the expression of $g^T$ and $N$ in terms of the physical degrees of freedom can be obtained from the two second class constraints (\ref{H_N-constraint}) and (\ref{H_P-constraint}), respectively.

The Lagrangian evaluated on the submanifold of constraints reduces to the kinetic terms plus surface terms
\begin{equation}
\label{Lagrangian}
L= \int dt d^3 x \,\pi^{ij}\,\partial_{t}g_{ij}  - E\,.
\end{equation}

We assume the following   flat asymptotic behaviour, as in GR:
\begin{eqnarray}
g_{ij}-\delta_{ij}=  \mathcal{O}(1/r) \,, \quad \partial g_{ij}=  \mathcal{O}(1/r^{2}) \,, \\ \pi^{ij}=\mathcal{O}(1/r^{2})\,,\quad \partial \pi^{ij}=\mathcal{O}(1/r^{3})\,, \\ 
N-1=  \mathcal{O}(1/r)\,,  \quad \partial N=  \mathcal{O}(1/r^{2})\,,
\\ N_i=  \mathcal{O}(1/r)\,, \quad \partial N_i=  \mathcal{O}(1/r^{2})\,.
\end{eqnarray}

Furthermore, we can fix, using the spacelike diffeomorphisms  $\mathcal{F}_{Diff}$, the coordinate condition \cite{ADM2008}
\begin{equation}
\label{coordinate_condition_1}
g_i= x^i + \left(\frac{1}{4 \Delta}\right) g^{T}_ {,i}  \,,
\end{equation}
or in differential form
%
%
\begin{equation}
\label{coordinate_condition_2}
g_{ij,jkk} -\frac{1}{4} g_{kj,kji}- \frac{1}{4} g_{jj,kki} =0 \,.
\end{equation}%

%

We will use an equivalent decomposition to (\ref{T+L}) but reorganized in a different way. We consider the following orthogonal  decomposition of $h_{ij}\equiv g_{ij}-\delta_{ij}$,
\begin{equation}
\label{h_{ij}-decomposition}
    h_{ij}=h_{ij}^{T\tau}+h_{ij}^{\tau}+\frac{1}{3}\delta_{ij}h,
\end{equation}
where $\delta_{ij}h_{ij}^{T\tau}=\delta_{ij}h_{ij}^{\tau}=0$, '$\tau$' means traceless with respect to $\delta_{ij}$. It follows that $h=h_{ii}$. By definition, $\partial_{i}h_{ij}^{T\tau}=0$, 'T' means transverse with respect to $\partial_{i}\,$,  and  $h^{\tau}_{ij}$ is defined in terms of a vector field $W_{j}$ 

\begin{equation}
   h_{ij}^{\tau}=\partial_{i}W_{j}+\partial_{j}W_{i}-\frac{2}{3}\delta_{ij}\partial_{k}W_{k}. 
\end{equation}
This decomposition exists and it is unique. It is analogous to the York decomposition, however it is not covariant. It can also be rewritten as
\begin{equation}
    \left(1+\frac{1}{3}h\right)h_{ij}=h_{ij}^{T\tau}+h_{ij}^{\tau}+\frac{1}{3}g_{ij}h.
\end{equation}
Comparing the decomposition (\ref{h_{ij}-decomposition}) with the T+L ADM decomposition (\ref{T+L}) we obtain
\begin{equation}
\label{comparacion_descomposiciones}
    h_{ij}^{T\tau}=h_{ij}^{TT}\,,\quad W_{i}=h_{i}-\frac{1}{4}\frac{\partial_{i}h^{T}}{\Delta}\,,\quad h=h_{ii}=h^{T}+2\partial_{i}h_{i}\,,
\end{equation}
where $h_{ij}^{TT}\,$, $h^{T}$ and $h_{i}$ are the T+L components of the ADM decomposition.
Using the  $\mathcal{F}_{Diff}$ symmetry we can always fix $W_{i}=0$. In fact, the gauge fixing condition is exactly (\ref{coordinate_condition_1}). We then have 
\begin{equation}
   h_{ij}=\frac{h_{ij}^{T\tau}}{1+\frac{1}{3}h}+\frac{1}{3}g_{ij}\frac{h}{1+\frac{1}{3}h}. 
\end{equation}

In order to analyse the reduction to the Hamiltonian density in terms of the physical degrees of freedom, the transverse traceless modes, we consider now the covariant decomposition introduced by  York \cite{york1973} 
\begin{equation}
 \pi^{ij}=\tilde{\pi}^{ijT\tau}+\tilde{\pi}^{ij\tau} +\frac{1}{3}g^{ij}\tilde{\pi}, 
\end{equation}
where $g_{ij}\tilde{\pi}^{ijT\tau}=g_{ij}\tilde{\pi}^{ij\tau}=0\,$, $\nabla_{i}\tilde{\pi}^{ijT\tau}=0$ and $\tilde{\pi}^{ij\tau}$ is defined in terms of a vector field $U^{j}$
\begin{equation}
    \tilde{\pi}^{ij\tau}=\nabla^{i}U^{j}+\nabla^{j}U^{i}-\frac{2}{3}g^{ij} \nabla_{k}U^{k}.
\end{equation}
It then follows from the constraints of the theory that 
\begin{equation}
    \pi^{ij}=\tilde{\pi}^{ijT\tau}. 
\end{equation}

We remark that $\tilde{\pi}^{ijT\tau}$ is a function of the $\pi^{ijTT}$ transverse-traceless ADM modes which are independent of the metric $g_{ij}$. One way of making explicit this point is to consider the sequence $\pi_{n}^{ij}\,$, $n=0,1,2,..., $  where $\pi_{0}^{ij} =\pi^{ijTT} $ and
\begin{equation}
    \partial_{i}\pi_{n+1}^{ij}+\Gamma_{ik}^{j}\pi_{n}^{ik}=0\, , \quad \delta_{ij}\pi_{n+1}^{ij}+h_{ij}\pi_{n}^{ij}=0.
\end{equation}
If this sequence is convergent, that is if there exists a fixed point of the procedure,  we obtain $\pi_{n}^{ij}\rightarrow \tilde{\pi}^{ijT\tau}$, which depends on the metric $g_{ij}$ and $\pi_{ij}^{TT}$ the transverse-traceless modes of the ADM decomposition.
We argue now that the Hamiltonian density in terms of physical degrees of freedom can be obtained in terms of $\tilde{\pi}^{ijT\tau}$ and $h_{ij}^{TT}$
\begin{equation}
    \pi^{ijTT}\, , \, h_{kl}^{TT}\quad \longleftrightarrow \quad \tilde{\pi}^{ijT\tau}\left(h_{mn}^{TT},\pi_{pq}^{TT} \right)\, , \,\,  h_{kl}^{TT}.
\end{equation}

The Lagrangian of the  Ho\v{r}ava-Lifshitz theory at the kinetic-conformal  point evaluated on the constrained submanifold is given by (\ref{Lagrangian}), where the surface terms $E$, (\ref{E_boundaries_terms}), arise from the Fr\'echet differentiability of the action and the expression of the Hamiltonian rewritten in terms of the constraints. 

We then have, using that $g_{ij}\pi^{ij}=0$ is a constraint, 
\begin{eqnarray}
    \pi^{ij}\partial_{t}h_{ij}&=&\pi^{ij}\partial_{t}\left( \frac{h_{ij}^{T\tau}}{1+\frac{1}{3}h}+\frac{1}{3}g_{ij}\frac{h}{1+\frac{1}{3}h}\right)\,\nonumber\\
   &=&\pi^{ij}\partial_{t}\left(\frac{h_{ij}^{T\tau}}{1+\frac{1}{3}h}\right)+\pi^{ij}\partial_{t}g_{ij} \frac{h}{3\left(1+\frac{1}{3}h\right)}\quad .
\end{eqnarray}
Also $\partial_{t}g_{ij}=\partial_{t}h_{ij}\,$, hence 
\begin{equation}
    \pi^{ij}\partial_{t}h_{ij}=\left[ \left( 1+\frac{1}{3}h\right)\tilde{\pi}^{ijT\tau}\right] \partial_{t}\left[ \frac{h_{ij}^{T\tau}}{1+\frac{1}{3}h}\right]\, ,
\end{equation}
where, in the gauge $W_{i}=0\,$, $1+\frac{1}{3}h=1+\frac{1}{2}h^{T}\,$, $h^{T}$ is the transverse component in the ADM decomposition and can be obtained from the constraints of the theory in terms of $h_{ij}^{TT}$. We then conclude that the canonical conjugate variables are 
\begin{equation}
    \frac{h_{ij}^{TT}}{1+\frac{1}{2}h^{T}}\, , \quad \left( 1+ \frac{1}{2}h^{T}\right)\tilde{\pi}^{ijT\tau}.
\end{equation}

The value of the Hamiltonian on the constrained submanifold is then given by the two surface terms in (\ref{E_boundaries_terms}). The Lagrangian  is
\begin{equation}
    L=\int dt d^{3}x
\left[(1+\frac{1}{2}h^{T})\tilde{\pi}^{ijT\tau} \right]\partial_{t}\left[\frac{h_{ij}^{TT}}{1+\frac{1}{2}h^{T}} \right]-E, 
\end{equation}
where $E$ is the surface term (\ref{E_boundaries_terms}). In contrast with the analysis in \cite{ADM2008}, in our argument there is no integration by parts.
\section{The gravitational fields in the wave zone}
\label{wavezone}
We define the wave zone, as in \cite{Arnowitt1961,ADM2008}, by the following  three conditions: First, $kr\gg1$ where $k$ is the wave number and $r$ is radial distance. This condition is the same that defines the far-zone in linear theories such as classical electrodynamics, and can be satisfied if the radial distance is far enough from the sources. In addition to the previous condition, for non-linear theories it is necessary to impose more restrictive conditions in order to ensure that the self-interaction does  not destroy the free propagation of the dynamical modes. Then as a second condition, we demand that the deviations  of the fields from "flat background"  are of the order of $\mathcal{O}(1/r)$, i.e.  $|g_{ij}-\delta_{ij}|\sim  |N-1|\sim   |N_{i}|= \mathcal{O} (A/r) \ll1$, where  $A(t,\theta,\phi)$  represent  generic functions of time, and angular coordinates such that $A$ and all its derivatives are bounded. 
The  third condition is
$|\partial g/\partial(kr)|^{2}\sim|\partial N/\partial(kr)|^{2}$
$\sim|\partial N_{i}/\partial(kr)|^{2}\ll|g-\delta|$, which is necessary to guarantee that the interaction of sub-leading order modes can not interfere to leading order $\mathcal{O}(1/r)$. 

The wave zone  in GR is a  region of space far away from the source where the TT modes of the metric, the physical degrees of freedom, propagate freely according to a linear wave equation. On the near zone the interaction terms are relevant and the propagation of the TT modes is not governed by the linear wave equation. The T mode of the metric and lapse $N$ are nontrivial in the wave zone and contribute to the gravitational energy and momentum.   Beyond the wave zone the dominant terms are the T mode and the lapse $N$ .

 %
 %
	%
	%
%
%

The solution of primary constraints (\ref{vinculo-momento}) implies that the propagating parts of the momenta in the wave zone behaves as
 \begin{eqnarray}
 \label{pi-estatico}
 \pi^{ijT}\sim \pi^{ijL}\sim \frac{B^{ij}}{r^{2}}+k\frac{\hat{A}^{ij}  e^{ikr}}{r^{2}},
 \end{eqnarray}	
 \begin{eqnarray}
 \label{piTT}
 \pi^{ijTT}\sim  \frac{B^{ij}}{r^{2}}+k\frac{\hat{A}^{ij}  e^{ikr}}{r} \,,
 \end{eqnarray}
we use $A$ y $B$ as generic tensorial functions of angles such  that  they and their   derivatives  are  bounded.

If we use (\ref{pi-estatico}) and (\ref{piTT}) in the secondary constrains (\ref{H_P-constraint}) and  (\ref{H_N-constraint}) and the transverse gauge $g_{ij,j}=0$, we obtain a coupled system of second-order elliptical partial differential equations for variables $g^{T}$ and $N$. We use here a different gauge condition than in the previous section. A more adequate one for the present analysis. This is admissible since the gravitational energy, the surface term $E$, obtained in the previous section is independent of the gauge condition. If we multiply (\ref{H_P-constraint}) by two and add the result to (\ref{H_N-constraint}) we decoupled the system and, since $\beta \neq \frac{\alpha}{2}$ for the values of $\beta$ and $\alpha$ determined from experimental data,  we can estimate $N$ and $g^{T}$ at low energies in the wave zone:
 \begin{equation}
 \label{N-estimate}
 N -1\sim g^{T}\sim\frac{B}{r} +\frac{\hat{A}  e^{ikr}}{r^{2}} \,,
 \end{equation} 
these fields have a nontrivial Newtonian part $\mathcal{O}(1/r)$. They are  zero in a linearized version of the theory.

In the complete version of the theory, where higher order derivative terms are included, there are oscillatory zero modes of the same order which have to be eliminated to obtain consistency of the theory. We do not have this problem in the low energy regime.
 
In the low energy regime, where only the contribution (\ref{potential1}) to the potential is considered, the term $\alpha a_{i}a^{i}$ breaks manifestly the relativistic symmetry. Its contribution to the present analysis can be determined directly from (\ref{N-estimate}). We obtain
 \begin{eqnarray}  
	a_{i}\lesssim \frac{B_{i}}{r^2}+  k \frac{\hat{A}_{i}  e^{ikr}}{r^{2}}, \\
	\label{a^{1}a_{i}}
	a_{i} a^{i}\lesssim \frac{B}{r^{4}}+ k^{2} \frac{\hat{A}  e^{ikr}}{r^{4}},\\
	\label{nabla^i a_{i}}
	\nabla^{i}a_{i} \lesssim \frac{B}{r^{3}}+k^{2} \frac{\hat{A}  e^{ikr}}{r^{2}}\,,
	\end{eqnarray} 
although terms including the vector $a_{i}$  are not involved in the field equations at order $\mathcal{O}\left(1/r\right)$,  they do contribute to the gravitational energy and to the Newtonian background.

From the transverse and longitudinal decomposition of the dynamical equations (\ref{gpunto}) and (\ref{pipunto}),  we obtain the canonical form of the wave equation
\begin{eqnarray}
  \partial_{t}g_{ij}^{TT}=2 \pi_{ij}^{TT}+\mathcal{O}(1/r^{2}),\\
  \partial_{t}\pi^{ijTT}=\frac{1}{2}\beta\Delta g_{ij}^{TT}+\mathcal{O}(1/r^{2}),
  \end{eqnarray}
equivalently, 
\begin{equation}
    \partial_{t}^{2}g^{ijTT}-\beta \Delta g^{ijTT}=0 +\mathcal{O}(1/r^{2}).
\end{equation}
Then to the leading  order $\mathcal{O}(1/r)$ the transverse traceless components of the spatial metric satisfy a wave equation   with speed of propagation  $\sqrt{\beta}$. Detection from gravitational waves arising from the merge of the neutron star binary system  \textbf{GW170817} \cite{Abbott2017} and its electromagnetic counterpart, ${\gamma}$-ray burst \textbf{GBR170817A} \cite{Abbott_2017GRB}, restrict the space of $\beta$-parame-ter to $| 1-\sqrt{\beta} |\leq 10^{-15}$ \cite{Abbott2019b}.
Then, if that is so,  the prediction on gravitational waves of Ho\v{r}ava-Lifshitz theory at the kinetic-conformal point, at low energies, is the same as in GR. 

\section{Discussion and conclusions}
\label{Conclusion}
We showed  that in Ho\v{r}ava-Lifshitz theory at the kinetic-conformal point,  in the low energy regime, a wave zone can be consistently defined. In it  the physical degrees of freedom, which reduce to the transverse traceless tensorial modes, satisfy a linear wave equation. 
The same one that arises from a linear perturbative approach \cite{Blas2011, BellorinRestuccia2018}, but unlike this there are Newtonian non-trivial contributions, that do not obstruct the free propagation (radiation) of the physical degrees of freedom. Among these contributions there are terms which manifestly break the relativistic symmetry. These terms which determine a different physical behavior, for example of the static, spherically symmetric solutions of the  Ho\v{r}ava-Lifshitz gravity theory compared to GR, do not contribute  to the free propagation of the physical degree of freedom to the  dominant order in the wave zone. However, they provide a relevant contribution to the gravitational energy and the gravitational momentum. 

The gravitational energy of the Ho\v{r}ava-Lifshitz gravity has been considered in \cite{Blas2011} from an asymptotic analysis, however the canonical reduction of the Hamiltonian to the physical degrees of freedom has not been analyzed.  In GR the reduction to a canonical formulation in terms of the TT modes can be achieved by a suitable gauge fixing condition. The main point is that the gauge fixing procedure for the Ho\v{r}ava-Lifshitz gravity reduces to a spacelike coordinate election, since only reparametrization on time is allowed. We obtained in section 3  the canonical reduction in terms of the true physical degrees of freedom, which at linearized level reduce to the TT modes of the metric. In general they are well defined functions of the TT modes of the metric.

We showed, using the $\mathcal{F}_{Diff}$ symmetry, that the Ho\v{r}ava-Lifshitz theory, at the kinetic-conformal point, can be canonically reduced to a Hamiltonian formulation in terms of the physical degrees of freedom,  in a particular coordinate frame. They reduce to the TT tensorial modes in the wave zone. 

In Ho\v{r}ava-Lifshitz theory there does not exist a universal (scale invariant) constant as the light velocity, however the energy-dependent coupling constant $\sqrt{\beta}$, in the renormalization flow from the UV regime to the IR point, should end up having a value very near or equal to the speed of light. In that case, although the Ho\v{r}ava-Lifshitz theory breaks the relativistic symmetry, the wave equation coincides with the relativistic one arising in GR. 

We expect  that the interaction terms with high spatial derivatives modify the wave equation by introducing linear high order spatial derivatives  in the wave zone, i.e $\Delta^{2}$ and $\Delta^{3}$ operators in the propagating equation. However, it is unknown if in this case the Newtonian background interacts in a non-trivial way with the propagation of the physical degrees of freedom. In particular, the resolution of the constraints is in this case a non-trivial problem.

\section{Aknowledgments}
J. Mestra-P\'aez acknowledge financial support from  Beca Doctorado Nacional 2019 CONICYT, Chile.   N° BECA: 21191442. J. M. P is supported by the projects ANT1956 and SEM18-02 of the University of Antofagasta, Chile. 

\bibliographystyle{elsarticle-num}
\bibliography{REF}

\end{document}